\documentclass{elsart}
\usepackage{natbib}
\begin{document}
\runauthor{Xu, Zhang and Qiao}
\begin{frontmatter}

\title{What if pulsars are born as strange stars?}

\author[Beijing]{R.X. Xu\thanksref{Bayreuth}}
\author[NASA]{Bing Zhang\thanksref{NRCRA}}
\author[Beijing]{G.J. Qiao}

\address[Beijing]{CAS-PKU Beijing Astrophysical Center
 and Astronomy Department, Peking University, Beijing 100871, China}
\address[NASA]{Laboratory of High Energy Astrophysics,
 NASA Golddard Space Flight Center, Greenbelt, MD 20771}
\thanks[Bayreuth]{Theoretical Physics IV, University of Bayreuth,
 95440 Bayreuth, Germany}
\thanks[NRCRA]{National Research Council Research Associate}

\begin{abstract}

The possibility and the implications of the idea, that pulsars are
born as strange stars, are explored. Strange stars are very likely to 
have atmospheres with typical mass of $\sim 5\times 10^{-15}M_\odot$ 
but bare polar caps almost throughout their lifetimes, if they are 
produced during supernova explosions.
A direct consequence of the bare polar cap is that the binding
energies of both positively and negatively charged particles at
the bare quark surface are nearly infinity, so that the vacuum
polar gap sparking scenario as proposed by Ruderman \& Sutherland
should operate above the cap, regardless of the sense of the magnetic 
pole with respect to the rotational pole. Heat can not accumulate
on the polar cap region due to the large thermal conductivity on
the bare quark surface. 
We test this ``bare polar cap strange star'' (BPCSS) idea with the 
present broad band emission data of pulsars, and propose several 
possible criteria to distinguish BPCSSs from neutron stars.

\vspace{5mm}

\end{abstract}


\end{frontmatter}

\section{Introduction}

Soon after the discovery of neutrons, the idea of neutron star
was proposed\cite{l32bz34}. The discoveries
of radio pulsars and some other accretion-powered X-ray
sources indicate that there do exist in nature certain
objects with mass $M\sim M_\odot$, and radius $R\sim
10$ km. They were commonly regarded as neutron stars
until the idea of strange star was proposed\cite{w84,afo86,hzr86}.
The basic idea of the strange star conjecture is that, 
strange quark matter (SQM), which is simply composed of an 
approximately equal portion of up, down, and strange quarks 
and a few electrons to balance the non-neutral charges, might 
be more stable than the normal nuclear matters.
According to the lattice quantum chromodynamics (QCD),
a new state of strong interaction matter, the so-called
quark-gluon plasma (QGP), will appear
when the temperature of the matter achieves as high as
$\sim150-200$ MeV or the density of the matter achieves
several times of $\rho_0$, where $\rho_0 
\sim 3\times10^{14}$g/cm$^3$ is the nuclear density. SQM 
is just one kind of such QGPs.
Though we can not tell whether SQM is the lowest state of
hadronic matter, it is found that the energy per
baryon of SQM is lower than that of the normal nuclear
matter for a rather wide range of QCD parameters\cite{b71,w84,fj84}.
Assuming SQM is absolutely stable at zero pressure,
strange stars, which almost completely consist of SQM,
could exist in nature\cite{afo86,hzr86}.
The assumption is {\em strong}, but not impossible.
Therefore many objects which were previously believed to be
neutron stars, including radio pulsars, might actually be
strange stars.
In fact, more and more strange star candidates have been
proposed recently in the literatures\cite{ldw95lbddv99,bom97,cdwl98,xqz99},
including both the accretion-powered and the
rotation-powered compact objects.

Therefore, we have to think over a question again, which
appeared to have been answered, ``What is the nature of pulsars?''
In their pioneering work, Alcock et al.\cite{afo86}
have suggested that radio pulsars might be strange stars.
However, they did not fully discuss the possible observational
consequences of such a strange star idea. Nor was this done by
the later authors. What if pulsars are born as strange stars? 
How well can a strange star model interpret the pulsar 
observational data as compared with a neutron star model? 
Is there any criterion based on the radiation properties that
can distinguish strange stars from neutron stars? These are some
interesting issues worth exploring, and are the main topics
of this paper. Suppose that strange stars are the
ground state of neutron stars, the conversion from neutron
stars to strange stars can occur in different stages of the
neutron stars' lifetimes. There are two main regimes discussed
in the literatures. One group of models suggests that such a 
phase conversion occurs in the very late stage of a neutron 
star's lifetime, after it accretes sufficient amount of material 
to make the core density exceed the critical density for phase
transition\cite{cd96,bd00}.
Another group of models, on which we mainly focus in this paper,
suggests that the conversion occurs in the very early stage of
a neutron star's life time, i.e., during or shortly after a
supernova explosion\cite{gen93dpl95aggs97}, so that the observed
radio pulsars might be strange stars.
In the conventional strange star models, a crust composed of
normal matter is usually assumed above a strange star's
bare quark surface\cite{afo86,kwwg95}, which tends to smear out the
possible differences between the emission of strange stars and of
neutron stars. An important issue raised in this paper which differs
from the conventional models is that strange stars are almost
bare, especially in the polar cap region,
if they are formed directly from supernova explosions.
Such bare polar cap strange stars (BPCSSs), can also well act 
as radio pulsars\cite{xq98,xqz99}.
In Sect.2, we will discuss that phase transition from nuclear 
matter to SQM may be an important mechanism to retain a successful 
supernova explosion, and the strange stars, if born as the products 
of the explosions, are very likely to have {\em very} thin crusts 
($\sim 10^{-15}M_\odot$) and bare 
polar caps almost throughout their lifetimes. The electrical and 
thermal properties of BPCSSs are discussed in Sect.3, focusing on 
some distinguishing properties as compared with neutron stars. In 
Sect.4, we test this BPCSS idea with the present available radio 
and high energy emission data of pulsars, and point out the strong 
and weak aspects of the idea. Finally we will propose some possible 
approaches to distinguish BPCSSs from neutron stars in Sect.5, and 
summarize the conclusions in Sect.6.

\section{Newborn strange stars}

\subsection{Supernova explosions: neutron stars or strange
stars?}

The theory of stellar structure and evolution has been proved
a great success in astrophysics. However, some challenges
still exist in understanding a star's life, such as the
core-collapse supernova paradigm, which begins with the
collapse of the iron core of an evolved massive star in
the end of its thermonuclear evolution (for a comprehensive 
review, see e.g. \cite{bet90}).
Whether a model yields a successful explosion
still challenges theorists and their numerical simulations.
It is currently believed that the prompt shock, which
results from the inner core's rebound after an implosion
has compressed the inner core to supranuclear density, can
not propagate directly outward and expel the entire envelope,
but may stall and turn into an accretion shock at a radius
of 100-200 km from the explosion center due to nuclear
dissociation and neutrino cooling.
It is a consensus now among different groups of supernova
studies, that a successful explosion model
requires a so-called ``shock reheating mechanism'' or
``delayed mechanism''\cite{wil85bw85}, in which the neutrinos
produced from the core are absorbed and/or scattered by the
materials in the stalled shock and the shock could be revived
to expel the envelope and give rise to a successful explosion.
Many current researches on core-collapse supernova mechanisms
are focused on the role of convection in the unstable regions
below or above the neutrinosphere, and it seems that the
supernova simulations without incorporating fluid instabilities
may fail to explode (see, e.g., the report by Wilson \& Mayle\cite{wm93}).
Generally, the simulations based on this mechanism give
too low an energy, e.g. 0.3-0.4 foe (1 foe $=10^{51}$ ergs),
to meet the observed energy of SN 1987A (at least 1.0 foe).
Recent two-dimensional simulations by Mezzacapa et al.\cite{mez98}
showed that the neutrino-driven convection is not adequate to give an
``optimistic'' 15$M_\odot$ supernova model, and the
simulated timescale for explosion is longer than what
is observed. Furthermore, these authors pointed out that
more realistic three-dimensional simulation may be more,
not less, difficult to obtain a successful explosion.

Therefore, the key criterion for a successful explosion
{\it and} its large enough energy should be the sufficient
neutrino energy deposition behind the stalled shock.
One possible way to enhance neutrino luminosity is
through phase transitions, such as the transition from
nuclear matter to two-flavor quark matter and from
two-flavor quark matter to strange quark matter
(SQM)\cite{gen93dpl95aggs97}.
After the core bounce, the central temperature and density
of a proto-neutron star with a typical radius $\sim 50~$km\cite{bl86}
might be high enough to induce
a QCD phase transition. A strange star will be finally formed
if SQM is absolutely stable. An important issue is that 
whether the central density could achieve the value for phase
transition. Unfortunately, this question is not easy to answer
due to a large uncertainty of the MIT bag constant $B$.
Schertler et al.\cite{sgt97} had included medium effects (by
effective quark masses), the influence of which can't be simulated
by taking into account the coupling constant ($\alpha_{\rm s}$) or
$B$ or strange quark mass ($m_{\rm s}$), in the study of SQM properties
and found that medium effects reduce significantly the SQM binding
energy (SQM could not be absolutely stable when $\alpha_{\rm s}>1$),
but can hardly change the mass and radius of strange stars if
they exist.
However, a recent investigation\cite{bl99} shows that
the phase transition from nuclear matter to quark matter 
can occur in proto-neutron stars as long as the bag constant
$B\leq$ 126 MeV fm$^{-3}$.
Various estimates of the bag constant\cite{bag_B} indicate
that the preferred value of $B$ lies in
the range of $60 {\rm MeV ~fm^{-3}}\leq B \leq 110 {\rm MeV
~fm^{-3}}$ (see, e.g., a review by Drago\cite{drago98}), which means 
that a phase transition is likely to happen. Furthermore, Drago
\& Tambini\cite{dt99} show that phase transition can occur at
densities slightly larger than $\rho_0$ in pre-supernova matter.

Suppose that the phase transition condition is met, the energy 
budget problem for supernova explosions can be then naturally 
resolved\cite{bh89}. Because each nucleon contributes
about 30 MeV energy during phase conversion\cite{cdwl98},
the total released phase transition energy
$E_{\rm pt}$ is then approximately
\begin{equation}
E_{\rm pt} = M c^2 \times {30 \over 931}
\sim 5.8 \times 10^{52} {M\over M_{\odot}}~~{\rm ergs,}
\end{equation}
where $M$ is the mass of the inner core.
The timescale to burn a protoneutron star to a protostrange star,
$\sim 10^{-4}$ s, is usually much smaller than that of neutrino
diffusion and thermal evolution, $\sim 0.5$ s\cite{bet90}, since
the actual combustion could be in the detonation mode\cite{detonation}
and the propagation velocity could be very probably close to
the speed of light. Thus, the neutrino luminosity $L$
caused by this conversion could be estimated as about
$E_{\rm pt}/0.5 \sim 10^{53}$ erg/s. In Wilson's 
computations\cite{wil85bw85},
the typical value of neutrino luminosity is $5\times 10^{52}$
erg/s, thus the total neutrino luminosity with the inclusion
of SQM phase transition should be $1.5 \times 10^{53}$
erg/s ({\it three} times that of Wilson value). The simulated
explosion energy should also be raised by a factor of three,
which is 0.9-1.2 foe, adequate to explain the observed
value from SN 1987A.

Another possible way to enhance neutrino luminosity is through
rapid rotation. A nascent neutron star can be rotating rapidly
shortly after the supernova explosion\cite{ls95},
though we have little knowledge about how fast it rotates
observationally and theoretically.
The inner core of a rapid rotating proto-neutron star is usually
more {\it massive} than that of a non-rotating one in order
to compress the inner core to about the same supranuclear
density at which the prompt shock wave occurs.
When the proto-neutron star de-leptonizes,
more neutrinos are emitted from a larger core, and the neutrino
luminosity behind the stalled shock could be enhanced to explode
the supernova successfully\footnote{
We note that a proto-neutron star with short rotation period
may generate strong magnetic field\cite{td93},
which could play an important role in energizing the supernova
shock if the magnetic reconnection or magnetohydrodynamic wave 
can significantly heat the neutrinosphere.
}.
Due to a larger centrifugal force
from rapid rotation, the central density of the core could
be much lower than the QCD phase transition density, so that
a neutron star could exist after the explosion.
However, such rapidly rotating neutron stars can not keep
being there for a long time because of the fine-tuning problem,
proposed by Glendenning\cite{glen97}. {\it Only those
neutron stars very close to the critical mass limit can rotate
rapidly} because they have the least radius and the greatest mass.
These stars have much higher central densities than the lighter
neutron stars. As these massive neutron stars spin down through
dipole electromagnetic and quadruple gravitational radiations,
the centrifugal force gets smaller and the central density could
be increased high enough for a QCD phase transition, and the
neutron stars will also inevitably be converted to strange
stars\footnote{
It is interesting to further investigate the corresponding
astrophysical appearence of such conversion (e.g., May such
kind of conversion act as the inner engine of the cosmic 
classical $\gamma$-ray bursts?).
}.

To sum up, the energy budget problem of the core-collapse
supernova paradigm seems to require extra neutrino sources,
and the phase transition from neutron matter to SQM is a
natural mechanism to cure the imperfection of the paradigm.
For the progenitors with negligible effect of rotation,
additional neutrino emission of phase transition could
sufficiently enhance the power of neutrino energy deposition
behind the stalled shock to retain a successful explosion
and its enough energy, with a strange star directly formed
after the explosion.
For the progenitors with rapidly rotating inner core, neutron
stars might be formed as semi-products after the
explosions, but
such neutron stars will be finally phase-converted to strange
stars when they spin down enough. Thus, it is plausible that
at least some known radio pulsars may be in fact strange stars 
rather than neutron stars. Firm conclusions could not be drawn
at present stage due to the large uncertainties involved in this 
problem. For example, most recently Mezzacappa et al.\cite{mezz00}
claimed that they can successfully simulate a 13$M_\odot$ supernova 
explosion with the exact Boltzmann neutrino transport, without 
invoking convection. Thus the question whether strange stars are
``obliged'' to be formed during the supernova explosions remains 
open. The aim of this paper is to explore the possible consequences 
of the assumption that strange stars are the final products, and such 
consequences could be tested by the observations and in turn, shed
light on the plausibility of the scenario itself.

\subsection{Newborn strange stars: crusts or not?}

There are two kinds of strange stars discussed in the literatures,
i.e., the strange stars with crusts composed of normal matter and
the bare strange stars (strange stars with bare quark surfaces).
Normal matter crusts above the bare quark surfaces of the strange
stars were raised by Alcock et al.\cite{afo86} in their pioneering paper.
However, they did not discuss the formation of such crusts, but
simply addressed that the universe is a ``dirty'' environment.
Glendenning \& Weber\cite{gw92} discussed the maximum crust mass,
$\Delta M\sim 10^{-5}M_\odot$ (later corrected to $\Delta M\sim
3\times 10^{-6}M_\odot$ by Huang \& Lu\cite{hl97}), a strange star can
sustain, mainly motivated by trying to reproduce pulsar glitches from
strange stars. But they did not discuss the formation of such crusts,
either. Usov\cite{usov97} considered a strange star with a crust (or an
atmosphere in his term) the mass of which is many orders of magnitude
lower than the maximum value [only $\Delta M\sim (10^{-20}-10^{-19})
M_\odot$], and studied the X-ray emission properties from such strange
stars. He supposed that a strange star is more likely of this type
shortly after its birth before the crust mass reaches the maximum.
In principle, a pure bare strange star is unlikely to exist in the
universe due to various accretion processes. The question is how thick
the crust (or atmosphere) could be in reality. In this paper, we will
show through some dimensional estimate that newborn strange stars 
are likely to have much thinner crusts than the maximum values and
that they will keep like this through their lifetimes, unless they 
encounter some special environments that make sufficient accretion 
possible. The main reasons for such a conclusion are the rapid 
rotation and the strong mass ejection as discussed below. We note, 
however, that thick crusts (close to maximum) should be formed in the 
strange stars in the accreting systems (such as the compact objects 
in X-ray binaries) or the systems that clearly have strong accretion 
history (such as the ``recycled'' millisecond pulsars).

\subsubsection{Rapid rotation}

A rapidly rotating newborn strange star can prevent accretion
from happening. {\it Only when a strange star rotates slowly
enough, could the accretion onto the surface be possible}.
There are three characteristic scale lengthes in describing the
accretion scenarios of magnetized compact stars.
The magnetospheric radius (or Alfven radius),
defined by equating the kinematic energy
density of free-fall particles with the magnetic energy
density, is
$
r_{\rm m}= ({R^6B^2\over \dot M \sqrt{2GM}})^{2/7}
\sim 3.2\times 10^{10} B_{12}^{4/7} R_6^{12/7}
M_1^{-1/7} \dot M_{10}^{-2/7}
$ cm, where $B=10^{12}B_{12}$ G is the surface magnetic
field of the star, $R=10^6R_6$ cm is the stellar radius,
$\dot M=10^{10}\dot M_{10}({\rm ~g~s^{-1}})$ is the accretion
rate, and $M_1=M/M_\odot$ is the stellar mass in unit of solar
mass.
The co-rotating radius, defined by the balance of gravitational
force and the centrifugal force, is
$r_{\rm c}=({GM\over 4\pi^2})^{1/3}P^{2/3} = 1.5\times 10^8
M_1^{1/3}P^{2/3}$ cm, where the rotation period $P$ is in unit
of second.
The third scale length is the radius of light cylinder, which
reads $r_{\rm l}={cP\over 2\pi} = 4.8\times 10^9 P$ cm.
Possible accretion of matter into a pulsar's magnetosphere
requires $r_{\rm m}<r_{\rm c}<r_{\rm l}$
(see, e.g., \cite{henri83} and \cite{bv91}).
The constraint of $r_{\rm c} < r_{\rm l}$ gives
\begin{equation}
P>3.1\times 10^{-5} M_1 ~{\rm s},
\end{equation}
which could be satisfied for all known pulsars. The requirement
that the Alfven radius is smaller than the corotating radius
($r_{\rm m} < r_{\rm c}$), however, is very tight, which reads
\begin{equation}
P > 3.2\times 10^3 B_{12}^{6/7} R_6^{18/7}
M_1^{-5/7}\dot M_{10}^{-3/7} ~{\rm s},
\label{accretion-2}
\end{equation}
since the Alfven radius could not get too low unless
the accretion rate is very high, e.g., $\dot M_{10} \gg 1$.
A similar discussion is presented. 
Note that in \cite{bv91} the maximum accretion rate (Eddington 
luminosity) is adopted.

Let's give a rough estimate on the accretion rate of a
solitary strange star.
Consider a strange star which is moving with a velocity 
$V$ through a gas with a density $\rho$.
The critical radius within which the dynamics of gas is
dominated by the stellar gravity is
$r_{\rm g} \sim {GM\over V^2}\sim 1.3\times 10^{12}M_1V_7^{-2}$,
where $V_7 = V/(10^7 {\rm cm/s})$. Thus according to Bondi
\& Hoyle\cite{bh44}, the accretion rate is roughly
\begin{equation}
\begin{array}{lll}
\dot M \sim 4\pi r_{\rm g}^2 \rho V
& \sim & 2.2 \times 10^8 M_1^2 V_7^{-3} \rho_{24} ~{\rm g~s^{-1}}\\
& \sim & 3.5 \times 10^{-18} M_1^2 V_7^{-3} \rho_{24}~M_\odot~{\rm yr^{-1}},
\end{array}
\end{equation}
where $\rho_{24}=\rho/(10^{-24}{\rm g~cm^{-3}})$.
By comparing a sample of pulsar proper motion data with
their Monte Carlo simulations, Hansen \& Phinney\cite{hp97}
found that the mean speed of pulsars at birth
is $V\sim (2.5 - 3.0)\times 10^7 {\rm cm~s^{-1}}$.
Diamond et al.\cite{djp95} investigated the local
interstellar medium (ISM) distribution with ROSAT all-sky EUV
survey, and found that the ISM mean density is $\sim 10^{-25}
-10^{-24}$ g cm$^{-3}$.
Thus the minimum rotation period for accretion of a typical
new born strange star in a typical ISM is $\sim 10^4$ s.
In supernova remnants, the medium density may be denser so
that $\rho\sim 10^{-22}$ g cm$^{-3}$. In this case, the
critical period for possible accretion is still $\sim 10^3$s.
Since the longest period of the present known pulsars is only
8.5s\cite{ymj99}, we conclude that
accretion is almost impossible for most isolated radio pulsars
in their lifetimes, unless they are born with much
less-than-average proper motion speed or they are in very
dense medium. Another reason against the possible accretion
is that when a pulsar is active, relativistic particles are
believed to flow out from the open field line region. Accretion
onto the polar cap is thus impossible with such an outflow.

\subsubsection{Mass ejection}

We have shown that accretion is almost impossible for a pulsar
after its birth. The question now is whether strong accretion
is possible during the supernova explosion.
It is known that a new born neutron star is very hot (the core
temperature $T_{\rm c} \sim 10^{11}$K), and the mass ejected
from such a hot neutron star during the first 10s is about $\sim
10^{-3} - 10^{-2} M_\odot$\cite{woosley}.
It is expected that a hot new born strange star should also have
such a high mass ejection rate, so that its surface is almost bare,
since the ejected mass is considerably larger than the
maximum crust mass of the strange star\cite{usov97,usov98}.
As discussed in section 2.1, the rate of energy release when forming
a strange star is $\sim 10^{53}$erg/s. Most of this energy
is carried away by neutrinos, but part of it will be converted to
electromagnetic radiation. A nuclei above the quark surface will feel
strong outward pressure by photons and neutrinos (note that the
Eddington luminosity is only $\sim 10^{38}$erg/s), therefore normal
nuclei can hardly remain being bound by the gravitational force
above a strange quark surface and will be pushed away. 
For a neutron star, the requirement of the hydrodynamical equilibrium
will make the core materials to self-adjust themselves to form a
stable structure, including a very heavy crust above the superfluid
neutron layer, and the backfalling supernova ejecta has little
contribution to such a crust. For a strange star, things are
quite different. Almost the whole iron core will be completely
converted to SQM, since SQM has the property to swallow all the
normal matter that directly contacts with it if Witten\cite{w84}'s
hypothesis is right.
Also the detonation flame may expel the outer part of a protoneutron
star to form a strange star with a bare quark surface\cite{detonation}
at the very beginning.
Thus a crust above the quark surface of a strange star then
solely depends on the fallback of the supernova ejecta.

There are a lot of uncertainties on the possible fallback of the
supernova ejecta. An important issue is whether a reverse shock can
be formed. It was argued that one needs to include a reverse shock to
fit the light curve of SN 1987A, and such a reverse shock might be due
to a sufficiently massive hydrogen envelope around the compact
star\cite{woo88}. If the formation of such a reverse shock is possible,
Chevalier and other authors\cite{che89} investigated the possibility
of the accretion with hypercritical (or ``super Eddington'')
accretion rate, and
found that the total accreted matter could be as high
as $10^{-3} M_\odot$ for a normal type II supernova, which is much higher
than the maximum crust mass a strange star can sustain. In such cases,
the strange star formed from the explosion should have a thick crust
with maximum mass. However, the Chevalier's scenario does not include
the influences of rotation and magnetic fields. When these effects
are taken into account, more likely, the backfalling materials will
form a disk\cite{mich88} rather than just directly falling back onto
the surface of the compact star. According to Michel\cite{mich91}, such a
disk is essential for pulsar magnetospheric electrodynamics. Some
models also invoked such a disk to interpret the emission behavior
of the anomalous X-ray pulsars\cite{chn99}.
Thus it is very likely that the mass which falls back to the pulsar
surface should be tiny compared to the maximum, as Usov\cite{usov97,usov98}
suggested.

Let us present a rough estimate of the mass and the thickness of the
crust formed due to the direct fallback. In principle, the materials
that can fall back onto the surface should be trapped by the
magnetosphere and be below the corotating radius $r_{\rm c}$.
When no more outflow pressure exists, all the materials within
the corotating ball will drop onto the surface if they are trapped
by the magnetosphere. The condition of magnetic trapping requires
the local magnetic energy density to be comparable of the kinetic
energy density of the ejecta, so that the trapping radius is
$r_{\rm t}=R B^{1/3} v^{-1/3} (4\pi\rho)^{-1/6}$,
where $v$ is the typical velocity of the ejected materials, and
$\rho$ is the mean density of the materials. We set $r_{\rm c}=
r_{\rm t}$ to roughly set the fallback condition, and get a rough
estimate of
$\rho \sim 7.0 \times 10^{-9}B_{12}^2 P^{-4}v_9^{-2} R_6^6 M_1^{-2}$
(about one day after the supernova explosion for $P=10$ms,
see Fig.28 in\cite{woo88}).
The total mass that deposits onto the surface
(i.e. the mass of the crust) is then $\Delta M=(4/3)\pi r_c^3
\rho$, which reads
\begin{equation}
\Delta M\sim 1.0\times 10^{17}B_{12}^2 P^{-2}v_9^{-2} R_6^6 M_1^{-1},
\end{equation}
or $\Delta M\sim 5\times 10^{-15}M_\odot$ for typical values
of $P =10$ms, $B_{12}=R_6=M_1=1$. Note that the typical velocity 
of the ejecta is adopted as $v=10^9 v_9$ cm/s, which is the typical 
velocity of the revived shock\cite{cf95}.
The adoption of a smaller velocity
will not influence our followup conclusion qualitatively.
Notice that $\Delta M$ obtained here is 10 orders of magnitude
smaller than the maximum $\Delta M$, but 5 orders of magnitude
larger than the ``atmosphere'' mass discussed by Usov\cite{usov97}.
We will call the structure formed by
the fallback materials a {\em massive atmosphere} rather than
a thin crust (following Usov\cite{usov97,usov98}).

We now estimate the depth of this atmosphere. The column
density of the atmosphere is $\sigma_{\rm a} \sim \Delta M /
(4\pi R^2)\sim 8.0 \times 10^3 B_{12}^2 P^{-2}v_9^{-2} R_6^4
M_1^{-1}$ g cm$^{-2}$. The bottom pressure of the atomsphere is
thus $p_{\rm b} = (GM\sigma_{\rm a}/ R^2)\sim 1.0 \times 10^{18}
B_{12}^2 v_9^{-2} R_6^2 P^{-2}$ dynes cm$^{-2}$.
According to the equation of state\cite{fow26}
\begin{equation}
p=1.8\times 10^{12}\rho^{5/3}~~{\rm dynes~cm^{-2}},
\label{eos}
\end{equation}
for a nonrelativistic, completely degenerate gas (since the density
is much smaller tha $10^6$ g cm$^{-3}$, see below), the bottom
density is about  $\rho_{\rm b}=5.8\times 10^5$ g cm$^{-3}$ for
$P=10$ms. Therefore  the scale height of a strange star atomsphere
is $h_0\sim \sigma_{\rm a}/ \rho_{\rm b}\sim 1.4\times 10^2$ cm for a 
10ms pulsar (thinner crust for longer period pulsars).
We note that such a atmosphere scale height is much larger than the
characteristic thickness of neutron star atomsphere ($\sim  0.01
- 1$ cm, see, e.g. \cite{zps96}). This is an essential feature.
It is known that a completely bare
strange star has very low emissivity in X-rays\cite{afo86}. We
emphasize here that {\em the existence of the massive atmosphere
described above makes a strange star to radiate thermal emission in a
similar way as does a neutron star} (except the polar caps,
see below). Another comment is that the ``atmosphere'' 
discussed here is conceptually different from the atmosphere of 
a neutron star, which is defined by the equilibrium of the 
gravitational energy and the thermal kinetic energy of the 
particles. Neutron star atmospheres are naturally formed even 
without accretion. The strange star ``atmospheres'' discussed in this 
paper, however, are formed due to the materials' fallback, and 
hence, are not subject to the definition above.

Shortly after a bare strange star is born, it is believed that a 
typical pulsar magnetosphere with Goldreich-Julian\cite{gj69} density 
will be formed slightly above the bare quark surface (typical height
$z_{\rm c}\sim 10^{-8}$cm, see Sect.3.1), mainly due to the pair
multiplication via $\gamma-B$, $\gamma-\gamma$\cite{xq98},
or $\gamma-E$\cite{usov98} processes. When such a magnetosphere
is formed, a space-charge-limited free flow\cite{arons83}
is then inevitable in the open field line regions, so that the
atmosphere materials above the polar caps, which are composed of
normal hadrons, will be pulled out. In neutron stars, such a flow
could persist throughout the pulsars' whole lifetimes since the 
thick crusts of the neutron stars can supply copious ions and 
electrons to be pulled out. In the case discussed here (a strange star
with an atmosphere), the particles available for this extracting is 
limited, which is only the part of atmosphere right above the polar 
caps. The time scale for pulling out this atmosphere is short. With a 
Goldreich-Julian flow, the extracting column number density flux 
is $F=(\Omega B/2\pi e)\sim 2.1\times 10^{21}B_{12}P^{-1}{\rm cm}
^{-2}{\rm s}^{-1}$. Given the atomic mass unit $u=1.66\times 
10^{-24}$ g, the extracting column mass density flux is then 
$F_{\rm m}\sim 2 u F$, therefore the extracting time scale is 
typically
\begin{equation}
\tau=\sigma_{\rm a}/F_{\rm m} \sim
3.6 \times 10^{-2} B_{12}P^{-1}v_9^{-2}R_6^4 M_1^{-1}~{\rm yr},
\label{tau}
\end{equation}
which is much shorter than a pulsar's lifetime. Furthermore, if the
timescale for the ejecta to fall back is longer than the timescale
of forming a Gorldreich-Julian magnetosphere, the space-charge-limited
flow from the polar cap region will effectively block the deposit of
the falling materials onto the polar cap area. We thus address
that {\em except for very young pulsars, the polar caps of the 
strange stars with massive atmospheres are likely to be completely 
bare when they act as pulsars}.

Note that in the above estimate (eq.[\ref{tau}]), we have assumed that
only the matter in the polar cap regions is stripped out. This can
be justified as follows. The magnetic field energy density at the pole
is $\epsilon_{\rm B} = {B^2\over 8\pi}\sim 4.0\times 10^{22}B_{12}^2$ 
ergs cm$^{-3}$. The energy density of the matter, which is essentially 
the electron degenerate energy density, could be estimated as
$\epsilon_{\rm e} = \epsilon_{\rm F} n_{\rm e}$, where $\epsilon_{\rm F}$ 
is the electron Fermi energy, $\epsilon_{\rm F}=2 \pi^4\hbar^4 c^2 n_e^2/
(e^2 B^2 m)\sim 1.03\times 10^{-63}B_{12}^{-2}n_e^2$\cite{r71}, and $m$ 
and ${n_{\rm e}}$ are the mass and number density of the electrons, 
respectively. For $n_{\rm e}$, we adopt the electron number density at 
the ``effective'' pulsar surface of a BPCSS (eq.[\ref{ne}] for $z=z_c$), 
i.e., $n_e \sim 10^{27} {\rm cm}^{-3}$. This gives $\epsilon_{\rm e} 
\sim 10^{18} {\rm ergs~ cm^{-3}}$, which is much smaller than 
$\epsilon_{\rm B}$. This prohibits the rapid penetration of the electrons 
across the magnetic field lines. However, the charged particles can also 
diffuse across the magnetic field lines due to the collisions between the
charged particles that gyrate around the lines, since a collision will 
alter a particle's velocity and make it to gyrate around another field line. 
The Fermi energy of the electrons in the case we are discussing is much 
higher than the energy interval between adjacent Landau levels of the
electrons, we thus adopt the classical descriptions to estimate the
diffusion rate. Folowing \cite{niu89}, the diffusion coefficient 
$D_{\rm c}$ is approximately given by the square of the mean distance 
travelled (which is of the order of the Larmor radius $\rho_{\rm L}={mvc 
\over eB}$) divided by the mean free flight time $\tau_{\rm F}$,
which is of the order of $ \tau_{\rm F}\sim {m^2 v^3 \over\pi e^4 n}$
(derived from eq.[2.55] or [2.65] of \cite{niu89}, assuming that light
elements dominate in the atomsphere). Thus the diffusion rate is 
$D_{\rm c}\sim \rho_{\rm L}^2/\tau_{\rm F} \sim {\pi c^2 e^2 n\over B^2 v}$.
Let's estimate the electron diffusion rate first. For the bottom density 
$\rho_{\rm b}\sim 5.8 \times10^5 {\rm g~cm^{-3}}$, the electron
number density is $n_{\rm b}\sim \rho_{\rm b}/(2u)\sim 10^{29} {\rm cm^{-3}}$,
the Fermi energy is $\sim 10^{-5}B_{12}^{-2} n_{\rm b,29}^2$ ergs,
and the electron velocity $v\sim c$, thus one has
\begin{equation}
D_{\rm c} \sim {\pi c e^2 n \over B^2} \sim
2.2\times 10^{-3} B_{12}^{-2}n_{b,29}~~{\rm cm^2 s^{-1}}.
\end{equation}
As particles diffuse, the scale height $h$ of the massive atomsphere
should be a function of the polar angle $\theta$, and also a function of
time $t$ before a steady state is achieved. After a certain period of 
time, the diffusion flow will be steady. Let's consider the following
equilibrium diffusion situation: at the polar cap boundary
($\theta=\theta_{\rm p}=1.45\times 10^{-2} B_{12} P^{-2/3}$),
$h=0$; at a much larger polar angle (e.g. the equator $\theta=\pi/2$), 
$h=h_0$.
Again using the equation-of-state (\ref{eos}), one gets
\begin{equation}
h(\theta) \sim {\sigma_{\rm a}(\theta) \over \rho_{\rm b}(\theta)}
\sim 3.0\times 10^{-18} n_{\rm b}(\theta)^{2/3}~~{\rm cm}.
\end{equation}
For a simple estimation of the upper limit, we consider a layer with 
height $h(\theta)$ and with a uniform density $n(\theta)\sim n_{\rm b}
(\theta)$ in the following discussions. In the steady diffusion regime, 
the diffuse rate  
\begin{equation}
I_{\rm df} = D_{\rm c} \cdot {{\rm d}n(\theta)\over R{\rm d}\theta}\cdot
2\pi R \sin\theta \cdot h(\theta)
\end{equation}
is a constant. Solving the above equation, one obtains 
\begin{equation}
3.1 \times 10^{21} h(\theta)^4 = I_{\rm df} \ln [{\tan (\theta/2)
\over\tan (\theta_{\rm p}/2)}],
\end{equation}
or $2.5 \times 10^{-49} n(\theta)^{8/3} = I_{\rm df} \ln [{\tan (\theta/2)
\over\tan (\theta_{\rm p}/2)}]$ for $n(\theta)-\theta$ relation.
Setting $n\sim 10^{29}$ cm$^{-3}$ for $\theta=\pi/2$, we get
an upper limit diffusion rate $I_{\rm df}\sim 10^{28}$ s$^{-1}$.
For smaller $n_{\rm b}$, one gets smaller $I_{\rm df}$.
On the other hand, the free flow rate with the Goldreich-Julian flux is
$I_{\rm SCLF}\sim \pi (R\theta_{\rm p})^2F\sim 10^{34}$ s$^{-1}$ for 
$P=10$ ms. This indicates that the diffusion is unimportant. As for the 
collision diffusion of the ions, the rate should be even smaller
as their moving velocity is much smaller due to their larger mass.
Furthermore, the total number of the nuclei in the atomsphere is 
$\Delta M/u \sim 10^{45}$, so that the timescale for the atomsphere height 
to change significantely is $10^{17}$ s ($> 10^9$ years). All these
indicate that the atomspheres can exist almost through the whole life of 
the pulsars, and that only the polar caps are likely bare.

Besides rapid rotation and high mass ejection discussed above,
there are some more reasons which support the picture we proposed. For
example, newborn strange stars usually have very high temperatures,
which can significantely reduce the Coulomb barrier, and can thus
result in fusion of nuclei and SQM by tunneling
effect\cite{kwwg95,usov97,usov98}. This is also in favor of the
formation of strange stars without thick crusts.

In conclusion, if a strange star is born as the product of a
supernova explosion, a thick crust is unlikely to be formed
via the direct fallback of the supernova ejecta. Only a very
thin crust, or a massive atmosphere, can exist above the bare
quark surface of the strange star, and the polar caps are
completely bare due to the space-charge-limited flows. Such a
situation usually remain unchanged through out the pulsar's
lifetime, since accretion is almost impossible due to rapid
spin of the pulsar. Hereafter we will define such strange 
stars as bare polar cap strange stars (BPCSSs). 

\section{Electric and thermal properties of the BPCSSs}

In this section, we will summarize the electric and thermal
properties of the BPCSSs discussed in the previous section. 
We will focus on some special properties of the BPCSSs as
compared with neutron stars.

\subsection{Electric properties}

Since a BPCSS could have bare quark surface in the polar cap 
region, it is important to investigate the electric characters 
near the quark surface.
As the strange quarks are more massive than the up and down quarks,
some electrons are required to keep the chemical equilibrium of
a strange star. This brings some interesting properties near the
bare quark surface of a BPCSS.
Since quark matter is bound through strong interaction, the density
change abruptly from $\sim 4\times 10^{14}$ g cm$^{-3}$ to nearly
zero in 1 fm at the surface, which is the typical length scale of
the strong interaction. The electrons, which are bound by the
electromagnetic interaction and are $\sim 10^3-10^4$ times less
denser than the quark materials, can spread out the quark surface
and be distributed in such a way that a strong outward static
electric field is formed. Adopting a simple Thomas-Fermi model,
one gets the Poisson's equation\cite{afo86}
\begin{equation}
{d^2 V\over dz^2} =
\left\{    \begin{array}{ll}
{4\alpha\over 3\pi}(V^3-V_{\rm q}^3) & z\leq 0,\\
{4\alpha\over 3\pi} V^3 & z > 0,
\end{array}     \right.
\end{equation}
where $z$ is the height above the quark surface, $\alpha$ is
the fine-structure constant, and $V_{\rm q}^3/(3\pi^2 \hbar^3 c^3)$ is the
quark charge density inside the quark surface.
A straightforward integration gives\cite{xq99}
\begin{equation}
{dV \over dz} =
\left\{    \begin{array}{ll}
-\sqrt{2\alpha\over 3\pi} \cdot \sqrt{V^4 - 4 V_{\rm q}^3 V + 3 V_{\rm q}^4} &
(z<0)\\
-\sqrt{2\alpha\over 3\pi} \cdot V^2 & (z>0)
\end{array}     \right.
\end{equation}
where the physical boundary conditions
$\{
z \rightarrow -\infty: V \rightarrow V_{\rm q}, dV/dz
\rightarrow 0;~~
z \rightarrow +\infty: V \rightarrow 0,   dV/dz
\rightarrow 0
\}$ have been adopted. The continuity of $V$ at $z=0$ requires
$V(z=0) = 3V_{\rm q}/4$, thus the solution for $z > 0$ finally leads to
\begin{equation}
V={3V_{\rm q}\over \sqrt{6\alpha\over\pi}V_{\rm q}z+4}~~
({\rm for}~z > 0).
\end{equation}
The electron charge density can be calculated as
$
V^3/(3\pi^2 \hbar^3 c^3)
$, therefore the number density of the electrons is
\begin{equation}
n_{\rm e}  =  {9V_{\rm q}^3\over \pi^2 
(\sqrt{6\alpha\over\pi}V_{\rm q}z+4)^3}
\sim {9.5 \times 10^{35} \over (1.2 z_{11} + 4)^3}
\;\; {\rm cm^{-3},}
\label{ne}
\end{equation}
and the electric field above the quark surface is finally
\begin{equation}
E = \sqrt{2\alpha\over 3\pi} \cdot
{9 V_{\rm q}^2 \over
    (\sqrt{6\alpha\over \pi} V_{\rm q} \cdot z + 4)^2}
\sim {7.2 \times 10^{18} \over (1.2 z_{11} + 4)^2}
\;\; {\rm V\;cm^{-1},}
\label{E}
\end{equation}
which is directed outward. In the above estimate, $V_{\rm q}\sim20$
MeV has been adopted, and $z_{11} = z/(10^{-11}$ cm).

It was believed that the strong outward electric field near the
quark surface has some implications on the properties of strange
stars. One implication is that a possible normal-matter crust
could be formed, which in this paper we claim to be thin.
Another issue is that, according to\cite{afo86}, ``a
rotating magnetized star with an exposed quark surface will
not supply the charged particles necessary to create a corotating
magnetosphere'', since ``the electric field induced by the
rotating magnetized star is small compared to the electric field
at the surface''. However, here we claim that this argument is
incomplete. A handy estimate from equation(\ref{E}) shows that,
although the electric field near the surface is about $5\times
10^{17}$ V cm$^{-1}$, the outward electric field decreases very
rapidly above the quark surface, and at $z\sim 10^{-8}$ cm, the
field gets down to $\sim 10^{11}$ V cm$^{-1}$, which is of the
order of the rotation-induced electric field for a typical
Goldreich-Julian\cite{gj69} magnetosphere. Here {\em we define the
critical height, $z_{\rm c}$, at which the strength of the
intrinsic electric field is equal to that of the rotation-induced 
field as the effective pulsar surface\cite{xqz99}, thus
a typical pulsar magnetosphere could be naturally formed above
this surface\cite{xq98}}.

\subsection{Thermal properties}

Investigations of the strange star cooling behavior have been performed 
by many authors. It has been argued that the cooling behavior of the 
young pulsars can act as a definite criterion to distinguish strange
stars from neutron stars, since strange stars may cool much faster
than neutron stars (e.g. \cite{piz91}). However, recent 
more complete analyses\cite{schaab97} on this issue indicate
that, direct Urca process could be also forbidden in strange stars
if the electron fraction of the SQM is relative low. Furthermore,
quarks may eventually form Cooper pairs, and such possible
superfluid behavior of the strange quark matter can also
substantially suppress the neutrino emissivities of various
processes. As a result, the surface temperature of the strange
stars with crusts (typically with maximum mass) should be more or
less similar to the neutron star surface temperature. Neutron stars
and strange stars with crusts are hence indistinguishable in their
cooling behaviors except for the first $\sim$30 years after their
births\cite{schaab97}. Present X-ray data from about 30
rotation-powered pulsars are consistent with the standard neutron
star cooling scenario\cite{schaab99}, and thus do not
contradict the idea of strange stars with thick crusts
(close to the maximum). Nonetheless, notwithstanding extensive
efforts, the study on this {\em global} 
cooling behavior of strange stars does not reach an agreement among
researchers, especially about the effect of the possible color
superconductivity in strange quark matter\cite{csc}.
In this paper, however, we suggest (see section 3.2.2) to focus 
on the {\em local} thermal properties of the polar caps, which may 
provide some distinguishing criteria for BPCSSs and neutron stars,
while to regard the global cooling behavior as an open issue to 
explore.

\subsubsection{Cooling of BPCSSs}

For the BPCSS picture we are discussing in this paper, since the
crust (atmosphere) is much thinner than the maximum, more explicit
work need to be done to explore the detailed cooling behavior as
compared with that of the strange stars with thick crusts.
This issue is beyond the scope of this paper. Nevertheless,
we expect that the cooling curves of the BPCSSs with atmospheres
may not differ too much from those of the strange stars with
thick crusts. The main reason is that the BPCSS discussed in our
case is not completely bare. A thick atmosphere ($\Delta M\sim
5\times 10^{-15}M_\odot$) above the surface of a bare strange star
makes that the strange star can dissipate the heat from the quark
core through thermal emission at the surface in a similar way that 
does a strange star with a thick crust (although the response
function could be quite different\cite{piz91}),
since the thickness of the layer fulfills
the optically-thick condition. We note that the atmosphere
described by Usov\cite{usov97,usov98} is much thinner [$\Delta M\sim
(10^{-19}-10^{-20})M_\odot$] than the atmosphere discussed in
this paper. In his case, the atmosphere is optically thin, and
the strange star has much higher hard X-ray emissivity than
a neutron star or a strange star with a thick crust.

The crust or atmosphere acts as a thermal insulator between
the hot quark core and the surface\cite{piz91}, thus
the thickness of the crust determines the time delay when
the surface and the core have a same temperature. A thinner
crust will drive the temperature dropping point on the
cooling curve to an even earlier epoch. For the case of pulsar,
the youngest pulsars with measured surface temperatures (or
upper limits) have the ages of several $10^3$ yr, much longer
than the dropping point on the cooling curves of the strange
stars with thick crusts ($\sim 30$ yr)\cite{schaab97},
thus thinner crusts on strange stars may not bring
inconsistency between the cooling theories and the
observational data.

\subsubsection{Polar cap heating}

Pulsar polar caps are believed to be hotter than the rest
of the surface due to the reheating by the bombardment of the
downward-flowing particles and their radiation. The
downward-flowing particles could be produced from the inner 
accelerator (either of vacuum type or of space-charge-limited 
flow [hereafter SCLF] type) or the outer gap, and the degree 
of polar cap heating is hence model-dependent. The vacuum gap 
model proposed by Ruderman \& Sutherland\cite{rs75} predicts
substantial polar cap heating, since the numbers of the
downward and upward particles are similar when the gap
breaks down due to the pair production avalanche (essentially
with the Goldreich-Julian density), so that the luminosity
deposited onto the polar cap is just the polar cap
luminosity brought by the primary particles, which reads
\begin{equation}
L_{\rm pc,v} = \gamma mc^2 \dot N_{\rm p} \simeq 1.1\times 10^{31}
\gamma_7 B_{12} P^{-2} ~{\rm ergs~s^{-1}},
\label{Lpcv}
\end{equation}
where $\gamma=10^7\gamma_7$ is the typical Lorentz factor of the
primary particles, and $\dot N_{\rm p}=c n_{_{\rm GJ}} \pi r_{\rm p}^2 = 1.4
\times 10^{30}R_6^3 B_{12} P^{-2}$ is the particle flow rate with
Goldreich-Julian density $n_{_{\rm GJ}}$. The SCLF 
models\cite{as79,arons83,hm98},
however, predict much less polar cap heating\cite{arons81,zh00a},
since only a small
fraction, $f=10^{-4}f_{-4}$, of the primary particles (see the
definition of the parameter $f$ in\cite{zh00a}, their eq.[63]) could be
reversed in the space-charge-limited electric field and be
accelerated back to the surface. The energy deposited onto the
surface in this model is then
\begin{equation}
L_{\rm pc,sclf}=f \gamma mc^2 \dot N_{\rm p} \sim 1.1\times 10^{27}
f_{-4}\gamma_7 B_{12} P^{-2} ~{\rm ergs~s^{-1}}.
\label{Lpcsclf}
\end{equation}
If outer gaps exist in some pulsars, the downward-accelerated
particles from these outer gaps may also hit the polar cap
eventually after loosing substential amount of their initial
energies, and deposit $\sim 5.9 P^{1/3}$ergs per particle when
they strike the surface\cite{wrhz98}. With Goldreich-Julian
density, the luminosity deposited onto the surface is
approximately
\begin{equation}
L_{\rm pc,og} \simeq 5.9 P^{1/3} \dot N_{\rm p} \simeq
8.2\times 10^{30} B_{12} P^{-5/3}
~{\rm ergs~s^{-1}}.
\label{Lpcog}
\end{equation}
Note that both the vacuum gap model and the outer gap model 
predict very strong polar cap heating, the luminosities of 
which are comparable to the total spin-down luminosity of 
the pulsars, which reads
\begin{equation}
L_{\rm sd} \simeq 9.7\times 10^{30}B_{12}^2 P^{-4}I_{45}
~{\rm ergs~s^{-1}}.
\end{equation}

The polar cap heating temperature can be estimated by 
$T_{\rm pc}=[L_{\rm pc}/(\sigma \pi r_{\rm p}^2)]^{1/4}$, where
$
r_{\rm p}=1.45\times 10^4P^{-1/2}
$cm is the polar cap radius, and
$\sigma=5.67\times 10^{-5}{\rm ergs\cdot cm^{-2}\cdot
K^{-4}\cdot s^{-1}}$ is the Stefan's constant. In the
curvature radiation controlled vacuum gap model (see
eqs.[\ref{Lpcv}],[\ref{Lgamma}]), the polar cap
temperature is $T_{\rm pc,v}=5.9\times 10^6
B_{12}^{3/14}P^{-3/14}$K. The space-charge-limited
flow model gives a lower temperature $T_{\rm pc,sclf}
=2.2\times 10^6 B_{12}^{1/14}P^{-1/14}$K (eq.[70] in\cite{zh00a}).
The outer gap model predicts
a medium temperature $T_{\rm pc,og}=3.9\times 10^6
B_{12}^{1/4}P^{-1/6}$K.

As the polar cap is hotter than the other part of a
pulsar's surface, heat may flow from the cap to the
surrounding area. If pulsars are neutron stars, such
heat flow is negligible, and the kinetic energy of the
backflowing particles and the energy of the
electromagnetic shower produced by these particles can
be almost completely converted into thermal energy and
be re-radiated back to the magnetosphere.
Let us give a rough estimate.
The coefficient of thermal conductivity for electron
transport in the neutron star surface\cite{jones78}
\begin{equation}
\kappa^{\rm NS} = 3.8\times 10^{14} \rho_5^{4/3}
~{\rm ergs}\cdot {\rm s}^{-1}\cdot {\rm cm}^{-1}\cdot
{\rm K}^{-1},
\label{kappa}
\end{equation}
where $\rho_5$ is the density in unit of $10^5$ g
cm$^{-3}$, is almost independent on the details of
the lattice. If we assume that the horizontal
temperature gradient in the crust of a neutron star is
roughly $\nabla T^{\rm NS}\sim T_{\rm pc}/ r_{\rm p}$,
and that the area of the heat flow is of the order of
$r_{\rm p}^2$, then the heat flow rate is roughly
\[
H^{\rm NS}\sim \kappa^{\rm NS} \nabla T^{\rm NS}
r_{\rm p}^2 \sim \kappa^{\rm NS} T_{\rm pc} {r_{\rm p}}
\]
\begin{equation}
\sim 5.5\times 10^{24} \rho_5^{4/3}T_6 P^{-1/2}
~{\rm ergs~s^{-1}},
\end{equation}
which is much smaller than $L_{\rm pc}$.
Therefore, the thermal conduction from the polar cap
to the surrounding area is unimportant\cite{jones78} for
pulsars being neutron stars.

However, if pulsars are BPCSSs, the electron number density
in the bare quark surface
$
n^{\rm BPCSS} = 1.5\times 10^{34} {\rm cm}^{-3},
$
(eq.[\ref{ne}] with $z_{11}=0$, see also\cite{xq99}),
is much larger than that of a neutron star,
$
n^{\rm NS} = 2.8\times 10^{28} \rho_5 {\rm cm}^{-3}
$, thus the thermal diffusion could be much effective.
The transport coefficients for degenerate quark matter due
to quark scattering had been calculated by Heiselberg \&
Pethick\cite{hp93}, from eq.[61] of which the thermal
conductivity of strange star
can be obtained,
\begin{equation}
\kappa^{\rm BPCSS} = 1.41\times 10^{22}({\alpha_{\rm s}\over 0.1})^{-1}
\rho_{15}^{2/3}~
~{\rm ergs}\cdot {\rm s}^{-1}\cdot {\rm cm}^{-1}\cdot
{\rm K}^{-1},
\label{kappabss}
\end{equation}
where $\alpha_{\rm s}$ is the coupling constant of strong
interaction, $\rho_{15}$ is the SQM density in
$10^{15}~{\rm g\cdot cm^{-3}}$. Thus we have
$\kappa^{\rm BPCSS}\sim 10^{22}~{\rm ergs}\cdot {\rm s}^{-1}
\cdot {\rm cm}^{-1}\cdot {\rm K}^{-1}$
for $\alpha_{\rm s}\sim 0.1, \rho_{15}\sim 1$.
The corresponding horizontal heat flow rate in
the bare quark surface is therefore of the order of
\begin{equation}
H^{\rm BPCSS}\sim 10^{32} ~{\rm ergs~s^{-1}},
\end{equation}
which is even larger than $L_{\rm pc}$. This means that
very likely, the heat deposited onto the polar caps
by the backflowing particles will be soon dissapated
to the other part of the BPCSS surface, so that no hot
polar cap could be sustained. 

It is worth noting that if the heat deposited onto the
BPCSS surface could be dissipated to the other regions at
the surface, this will add an additional full surface
thermal component besides the cooling component. The
temperature of this component can be estimated as
$T_2=[L_{\rm pc}/(\sigma 4\pi R^2)]^{1/4}$. For the vacuum
gap case, this is
$T_2\simeq 5.0\times 10^5 B_{12}^{3/14}P^{-13/28}$K. The
inclusion of this component does not contradict the
cooling observational data for the normal pulsars\cite{schaab99}.
For the millisecond pulsars, since
these pulsars no longer appear as BPCSSs, such a
relationship no longer holds. We note that the extra full
surface thermal component is also expected in the internal
heating model\cite{schaab99} and the outer gap
model\cite{wrhz98,cz99}.

\section{Testing BPCSS idea with pulsar data}

Both neutron stars and strange stars have been proposed as the
nature of pulsars. However, almost all the previous researches
have invoked thick crusts above the bare quark surfaces, which
tend to smear out the information from the strange quark matter
itself. As a 
result, distinguishing strange stars from neutron stars is a
difficult task. One has to appeal to some other criteria to
do the discriminations. These include equation-of-state\cite{ldw95lbddv99}, 
cooling behavior\cite{piz91,schaab97}, 
dynamically damping effect\cite{wl84}, 
minimum rotation period (e.g. whether there exist sub-millisecond
pulsars)\cite{madsen92}, the vibratory modes\cite{brod98}, and so on.
Some of these criteria, e.g., cooling behavior,
suffer large uncertainties. We have shown that if
pulsars are born as strange stars, they will very likely appear as
BPCSSs. The exposure of the bare quark surface at the polar caps 
makes it possible that the information from the surface can be 
directly transmitted out, 
which may influence pulsar emission behavior to some extent. This 
opens a new window to distinguish strange stars from neutron
stars according to their {\em different emission behaviors}. In 
this section we will discuss the consequences of the bare polar 
caps and test the BPCSS idea with the present available pulsar 
data. 

\subsection{Consequences of the bare polar caps}

Two direct consequences follow naturally from the BPCSS picture.

The first consequence is that the inner accelerator in the polar 
cap region is vacuum-like, as proposed by Ruderman \& 
Sutherland\cite{rs75}, since the binding energy at the surface is 
almost infinity for both positive and negative charges.
There are two sub-types of pulsar inner gap models, according to the
boundary condition at the surface, i.e., the vacuum gap model\cite{rs75} 
and the SCLF model\cite{as79,arons83,hm98}. Though both models share 
some common features, they are different in some other aspects (for
a comparison between the two models, see \cite{zhm00,zh00b}).
If pulsar are neutron stars or strange stars with thick crusts, 
binding energy calculations favor the SCLF scheme, both for the case
of ${\bf \Omega}\cdot{\bf B}<0$ and ${\bf \Omega}\cdot{\bf B}>0$ (see 
a brief review in \cite{um95,xqz99}). However, if pulsars are BPCSSs, 
vacuum gaps are preferred. This is obvious for the case of ${\bf 
\Omega}\cdot{\bf B}<0$, i.e., anti-parallel rotators, because in 
this case positive charged $u$ quarks, which are expected to flow 
out, are definitely impossible to be pulled out since they are bound 
by strong interaction. For the case of parallel rotators (i.e. 
${\bf \Omega}\cdot{\bf B}>0$), electrons, which are bound
electromagnetically, might be pulled out, contingent upon
the competition between the intrinsic electric field at the BPCSS
surface (eq.[\ref{E}]) and the rotation-induced electric field.
With (\ref{E}), we see that only those electrons above the
height $z_{\rm c}$ could be pulled out. Thus a vacuum gap
could be formed above the ``effective'' BPCSS surface defined
by $z_{\rm c}$. As we have discussed in Sect.2.2.2, electrons 
are forbidden to cross field lines due to the strong magnetic 
field energy density, thus only electrons above $z_{\rm c}$ in the polar
cap region could be stripped out. The time scale for stripping
these electrons is about $10^{-5}$s\cite{xq99}, which is quite 
small. A steady SCLF accelerator is therefore not possible.
Although in the very beginning of the growth of gap it is not 
completely vacuum, the gap will be eventually evacuated
when the flow ceases. Thus the accelerators for the parallel
rotators are also vacuum-like.

The basic picture of the vacuum gaps formed above the polar
caps of pulsars have been delineated explicitly by Ruderman
\& Sutherland\cite{rs75}.
In this picture, a vacuum gap is formed right above the
surface of the star. Primary particles are accelerated to 
extremely relativistic energies and emit $\gamma$-rays via 
curvature radiation\cite{rs75} or inverse Compton scattering 
with the thermal photons near the surface\cite{xia85,zq9697}.
These $\gamma$-rays are materialized in the
strong magnetic fields via the $\gamma-B$ process, and the
secondary pairs screen out the parallel electric fields
so that the vacuum gap is limited at a certain height (the
gap height). 
A main feature of the vacuum gap which distinguishes from a SCLF
gap is the periodic breakdown of the gaps, which produces sparks 
and secondary plasma clumps ejected into the outer magnetosphere.
Since the curvature radius of magnetic field lines is smaller near 
the polar cap edge, the sparks tend to take place in a ring-like 
region near and within the polar cap edge. The secondary plasma 
produced in the sparks hence form some plasma columns or mini-tubes. 
Each spark forms a clump, and these plasma clouds are ejected 
sporadically from the gap and are separated from each other 
spatially. Due to the ${\bf E\times B}$ drifting, the sparks are
expected to rotate around the magnetic pole with a certain speed, 
and recent long term observations have revealed that the drifting 
subpulses observed in some pulsars clearly match the Ruderman \& 
Sutherland's\cite{rs75} prediction\cite{drifting}. 
We hope to emphasize that, although the vacuum gap model
was first proposed to operate on neutron stars,
later binding energy calculations indicate that such
gaps can be hardly rooted to neutron stars\cite{um95,xqz99}). 
The inner gap rooted to a neutron star or a strange star with 
thick crust has to be modified into other forms, e.g., the 
completely free-flow without any binding\cite{arons83} or the 
flow with partial binding\cite{cr80,um95}. Since none of these 
models can reproduce the exact drifting features predicted by 
the vacuum model, the recent observational results\cite{drifting}
could be regarded as an important support to the BPCSS 
scenario\cite{xqz99}.

The second direct consequence of the BPCSS scenario is
that hot polar caps can not be formed due to the large thermal 
conductivity as discussed in Sect.3.2.2. We will discuss more
about the implications of this feature for pulsar X-ray 
emission theories in Sect.4.4.

More than 1000 radio pulsars have been detected so far. Among 
them, 35 are detected as X-ray sources, 11 are X-ray pulsars,
and 8 are $\gamma$-ray pulsars\cite{radioPSRs,bt97,thhu97}.
A wealth of broad band emission data have been accumulated. 
In the following, we will test the BPCSS idea with the pulsar 
broad band emission data, focusing on the possible difference 
between the neutron star scenario and the BPCSS scenario. 

\subsection{Radio emission}

Pulsar radio emission data are abundant compared with most of
the other astrophysical objects. However, theories lag 
observations a big phase. More than 10 models have appeared in 
the literatures\cite{melrose95}, and they differ from each other
on many aspects. Perhaps the only consensus among these models is 
that pair production is the essential condition. Since both vacuum 
gaps and SCLF gaps can produce secondary pairs, pulsars will emit
radio emission as long as one of these inner accelerators 
operate\cite{zhm00}. To distinguish BPCSSs from neutron stars,
one needs to seek observational properties which characterize vacuum
gaps rather than SCLF gaps, or vice versa.

As discussed above, the clearly drifting subpulse patterns 
discovered recently\cite{drifting} seem to be a support to the 
BPCSS scenario rather than the neutron star scenario. Definite 
conclusion can not be drawn until it is proved that (1) SCLF 
models or any other modified vacuum gap models definitely can 
not reproduce the right drifting rate as observed, and that (2) 
there is no way to solve the binding energy problem within the 
neutron star scenario. At present, it seems that BPCSS scenario 
is the only way to revive the vacuum gap model, and to interpret 
the subpulse drifting rate.

Some observational features also directly infer the ``sparking''
behavior from the pulsar inner gaps. Individual pulses are often
composed of one or more sub-pulses, and some of these subpulses
drift regularly. Pulsar micro-structures have fine structures
of the order of $10^{-5}-10^{-6}$ s, which is the breakdown 
timescale of a typical Ruderman-Sutherland gap.

Besides these direct effects, it seems that vacuum gaps have 
advantages to interpret some other radio emission data than 
SCLF gaps. Generally speaking, self-absorption of the radio 
emission limits the incoherent brightness temperature to a 
level much lower than what is observed. To interpret the 
extremely high brightness temperature observed from pulsars, 
certain coherent mechanisms are believed to play the role. 
Various methods have constrained the height of radio emission 
to be well within the light cylinder\cite{site}. However, the 
strongest electromagnetic instabilities, i.e., the maser 
instabilities (which do not depend on the type of the inner
gaps), occur at altitudes close to the light cylinder\cite{lbm99}, 
which is not favored by the empirical laws of pulsar radio 
emission uncovered by Rankin\cite{rankin8390}. Most low-altitude 
radio emission theories, on the other hand, require that
inner accelerators should display certain ``oscilation'' 
behaviors which resemble the quasi-periodical breakdown of the 
vacuum gaps. 

For example, the inverse Compton scattering model proposed 
by Qiao \& Lin\cite{ql98} attributes pulsar radio emission to
the coherent inverse Compton emission of the secondary particles.
This model can reproduce radio pulsar phenomenology well, 
including one core and two conal emission components found by 
Rankin\cite{rankin8390}; the linear and circular polarization 
features\cite{xqh97,xlhq00} and the frequency evolution of the 
pulse profiles\cite{qlzh00}. The basic ingredient of this model 
is a vacuum gap, the periodic breakdown of which can 
naturally excite the low-frequency electromagnetic wave, which 
is the target of the inverse Compton scattering of the particles. 
Furthermore, since the pair plasma ejected from the gap is 
confined to plasma columns in strong fields, the outflowing 
plasma should be highly inhomogenous in space, and the density 
between the miniflux tubes could be sufficiently low, which allows 
the low-frequency radio wave to propagate as if in vacuum.

Another example is the spark-associated soliton model recently 
proposed by Gil et al.\cite{gil00}. According to the authors, the 
plasma instability invoked to interpret pulsar radio emission is 
the only known low-altitude instability. Such over-taking two
stream instability requires pair plasma to be ejected in clumps,
and the sparks produced from a vacuum gap can naturally provide
such spacially separated plasma clumps.

We are not saying that SCLF model can not interpret pulsar radio
emission, though. In fact, in the neutron star scenario, a SCLF
gap is preferred from the binding energy calculations, and some
pulsars, e.g., the newly discovered 8.5s pulsar 
PSR J2144-3933\cite{ymj99}, are preferably interpreted by the 
SCLF model\cite{zhm00}. Even in the strange star scenario, 
millisecond pulsars, which show similar emission behavior as
normal pulsars, should also have SCLF accelerators since we
expect thick crusts above the quark surfaces formed from their
``recycling'' history. What we hope to address is that, certain
pulsar emission features (e.g. drifting subpulses) do not favor
the SCLF model, and these features are not observed from the
millisecond pulsars and the 8.5s pulsar. Thus in order to 
interpret all the pulsar radio emission features (especially 
the drifting subpulse phenomenon which is popular among conal
emission pulsars) within the neutron star scenario, either the 
SCLF model should be such modified to include certain periodic 
oscillation behaviors, or some fundamental progress is made to
solve the binding energy problem faced by the vacuum gap model.
At present stage, neither of these two possibilities are 
promising, therefore at least for some pulsars, BPCSS scenario 
is favored.

\subsection{Gamma-ray emission}

In contrast to the radio emission theories, pulsar $\gamma$-ray 
emission theories are usually grouped into only two types, the 
polar cap cascade models\cite{dh96,sdm95,zh00a} and the outer gap
models\cite{chr86,romani96,cz99} (for a comparison between the
two models, see e.g. \cite{zh00a}). As outer gaps are much far 
away from the surface (above the null charge surface), different 
polar cap properties will not bring differences in the outer gaps,
thus $\gamma$-ray data can not be used to differentiate the BPCSS 
scenario and the neutron star scenario if $\gamma$-rays are of 
outer gap origin.

If pulsar $\gamma$-rays are of polar cap origin, however, 
different types of the inner accelerators will bring differences
in the $\gamma$-ray emission properties. Contrary to radio emission, 
$\gamma$-ray emission data seem to favor the SCLF picture rather 
than the vacuum gap picture. But as we will show below, the 
BPCSS picture, i.e. the vacuum gap picture, can not be 
completely ruled out.

The basic $\gamma$-ray emission properties of the known 
$\gamma$-ray pulsars include a luminosity law $L_\gamma \propto
(L_{\rm sd})^{1/2}$\cite{thhu97}, and wide separations between 
the double peaks observed in several pulsars. Within the curvature
radiation controlled SCLF model\cite{hm98}, the luminosity law is
well obeyed since the typical Lorentz factor of the primary 
particles does not sensitively depend on pulsar parameters
($P$, $B_p$)\cite{zh00a}. The wide separations of the double peaks
could be interpreted by the extended polar cap scenario\cite{dh96},
since the SCLF accelerator may be lifted to a higher altitude due 
to the anisotropy of the inverse Compton scattering of the upwards
versus downward primaries\cite{hm98}.

If the inner accelerator is of vacuum-type, the gap should be 
formed right above the surface. This does not favor the 
interpretation of the $\gamma$-ray emission. First, it is
possible that there exist some strong multipole magnetic field
components near the surface (these components have been long
assumed\cite{rs75}), which tend to lower the gap height and 
limit the achievable $\gamma$-ray luminosity (essentially the
polar cap luminosity). However, if we assume that {\em the 
near surface magnetic field configuration is dominated by the 
dipolar component} for the known $\gamma$-ray pulsars, the
$L_\gamma \propto (L_{\rm sd})^{1/2}$ luminosity law can be
retained. For a curvature radiation controlled vacuum gap, with
pure dipolar field, the total voltage across the gap is
$\Delta V = 2.1\times 10^{13} P^{1/7}B_{12}^{-1/7}$ V, and the
typical Lorentz factor of the particles is $\gamma = 4.0\times
10^7 P^{1/7} B_{12}^{-1/7}$. With (\ref{Lpcv}), we get
\begin{equation}
L_\gamma \simeq L_{\rm pc,v} = 4.4\times 10^{31} B_{12}^{6/7}
P^{-13/7} ~{\rm ergs~s^{-1}},
\label{Lgamma}
\end{equation}
which is similar to the result of the SCLF model with no 
electric field saturation [eq.(59) in\cite{zh00a}].
This is understandable, since when the SCLF accelerator is not 
saturated, the $\Delta V-h$ law is also approximately quadratic 
($\Delta V \propto h^2$), just as the vacuum gap model\cite{zhm00}.
Equation (\ref{Lgamma}) will also give a similar diagram 
as Fig.3 in\cite{zh00a}, which approximately
reproduces the $L_\gamma \propto (L_{\rm sd})^{1/2}$
feature as reported by the observations\cite{thhu97}.

Another drawback of the BPCSS scenario is that, since vacuum 
gaps are formed on the surfaces, very small inclination angles 
between the magnetic axis and the rotational axis are required 
to account for the observed widely separated double 
$\gamma$-ray peaks. The detectability of the 
$\gamma$-ray pulsars is also lowered. Some statistic 
studies show that such small inclination angles are not 
contradictory to the number of the $\gamma$-ray pulsars 
presently detected\cite{ds94}, but one needs to answer such
questions like ``Why young pulsars are born with very small
inclination angles?''. Nonetheless, keeping in mind that outer
gaps are another alternative of $\gamma$-ray emission site,
the BPCSS scenario does not strongly contradict the pulsar
$\gamma$-ray emission data.

\subsection{X-ray emission}

In the neutron star scenario, X-ray emission of the 
spin-powered pulsars could in principle have three components, 
i.e., a non-thermal component with typical power-law spectrum, 
a thermal component from the full neutron star surface, and a 
hotter thermal component at the polar cap. The non-thermal
component has been observed in many pulsars; the full-surface
thermal component has been observed in 4 pulsars\cite{thermal1}; 
and the evidence for hot polar cap emission is only strong for 
the millisecond pulsar PSR J0437-4715\cite{thermal2} (for a 
brief review, see e.g. \cite{zh00a}). A luminosity law, $L_{\rm x}
\sim 10^{-3} L_{\rm sd}$, was found\cite{bt97}.

These data again could be interpreted either by the outer gap
model\cite{cz99} or the full polar cap cascade model\cite{zh00a}.
The outer gap model, which attributes the non-thermal X-ray emission
to the synchrotron radiation of downward cascade particles\cite{cz99},
will make no difference between the BPCSS scenario and the neutron
star scenario. The polar cap model, which attributes the non-thermal
X-ray emission to the inverse Compton scattering of the upward cascade
particles\cite{zh00a}, may give different predictions between the
two different scenarios. However, if we again assume dipolar-dominated
configuration for the young pulsars, the non-thermal X-ray luminosities
will not differ too much from the one predicted in the SCLF 
model\cite{zh00a}, since both models have similar polar cap 
luminosities (eq.[\ref{Lgamma}]).

Thermal X-ray emission, especially that from the polar caps, may be a 
tool to differentiate between the two scenarios. In the neutron star
scenario, only the SCLF model predicts a cool polar cap; both the vacuum
gap model and the outer gap model predict a much hotter polar cap than
observed (see Sect.3.2.2). To avoid such an inconsistency, the outer gap 
models have invoked an assumed pair blanket slightly above the surface, 
which can reflect the hard thermal emission from the polar cap back to 
the neutron star surface\cite{wrhz98,cz99}. An important consequence
of the BPCSS scenario is that no hot polar cap could be formed 
(Sect.3.2.2), which can naturally amend the hot polar cap problem 
encountered by both the outer gap model and the vacuum gap model. 
Evidence of a hot polar cap from the millisecond pulsar PSR J0437-4715 
was noticed (e.g. \cite{thermal2}), but this poses no objection to the 
BPCSS scenario since millisecond pulsars are believed to have thick crusts.

\subsection{Other issues}

Besides the broad band emission data, some other pulsar phenomenology may 
also shed light on the discrimination between the two scenarios. 

Pulsar ``glitches'' have been observed in some young pulsars. The
giant glitches observed from the Vela pulsar is regarded as a strong
evidence against the strange star model, even for the strange stars
with the thickest crusts\cite{alpar87}. The BPCSS scenario is even
disfavored. We may think that glitching pulsars might be neutron stars,
but firm conclusion still can not be drawn, since some efforts have 
been made to produce strong glitches in strange stars.
Benvenuto \& Horvath\cite{bh90} presented a
calculation through the introduction of the effects of the 
hypothetical few-quark bound state (quark-alpha). An important 
feature emerging from their calculation is that strange stars 
may become shell-structured when quark-alpha particles are
introduced. Glitches may arise from these shells with a similar
behavior in the shells of the neutron stars. They further suggest that 
the pulsars' post-glitch behavior observed should be attributed mainly 
to the de-coupling and re-coupling of the fluid-quark-alpha layer.
However, they note that some important problems remain unsolved. 
More studies are necessary before firm conclusions can be drawn.

Recently, Madsen\cite{madsen99} argued that bare strange stars can
be ruled out as the candidate of the very fast millisecond pulsars,
according to the dependence of the rotational mode instabilities on
the thermal behavior of the strange stars. This is an interesting
test, but is unfortunately not applicable to the case discussed here,
since in our BPCSS scenario, the millisecond pulsars have thick 
crusts due to their accretion history.

\section{Distinguishing BPCSSs from neutron stars}

In this paper, we are trying to open a new window to distinguish
strange stars from neutron stars using the exterior properties
of pulsars which are of {\em magnetospheric-origin}. 
As we have shown, the idea of BPCSS
raises some interesting distincting properties between the two 
scenarios. Although present data are not sufficient to draw definite
conclusions, the discrimination between the two scenarios is expected
as data accumulate. Here we summarize some distinguishing criteria
which may act as a guide for future observations.

1. Clearly drifting patterns from the ``antipulsars''. As discussed
in\cite{xqz99} and above, drifting pulsars give a strong support to 
the BPCSS idea. The support is more prevailing if the pulsar is a 
parallel rotator (${\bf \Omega\cdot B}>0$), since in this case, 
vacuum gaps are even less impossible to be formed above a neutron 
star's surface and it is referred to ``antipulsar'' in Ruderman \& 
Sutherland\cite{rs75}'s term. Unfortunately, the sense of the magnetic 
pole is indistinguishable from the polarization data. Seeking other 
methods to tell the sense of the magnetic pole is essential.
The premise of this criterion is that the SCLF model can not be
developed to manipulate the ``sparking'' behavior and to predict
correct drifting rate.

2. Polar cap temperature. Discussions in Sect.3.2.2 suggest
that most probably, BPCSSs might not have hot polar caps. This
raises the possibility to distinguish strange stars from neutron
stars using the {\em local} thermal behavior rather than the
{\em global} thermal behavior (which is reflected in their
cooling curves and is subject to many microphysics processes
poorly known). A very hot polar cap on a normal isolated
pulsar (no accretion history) will be a strong evidence against the 
BPCSS idea. Notice that the lack of polar cap temperature as high 
as several $10^6$K can not rule out the neutron star model since 
the SCLF models predict much lower polar cap temperatures [see e.g. 
eqs.[70],[73] in\cite{zh00a}]. However, if there is completely no 
evidence for hot polar cap emission, then the BPCSS idea is justified.

3. Line features. The different composition of the surfaces 
of neutron stars and BPCSSs may provide another interesting
feature to differentiate between the two scenario. Neutron star
surface is copious of positive ions. Within the SCLF picture, 
these ions should be pulled out from the surface and fill the
magnetosphere in the open field line region for those pulsars
with ${\bf \Omega\cdot B}<0$.
For BPCSSs, however, nothing can be pulled out from the
quark surface, and the magnetosphere is just filled with the
electron-positron pairs.
The different compositions in the open field line region of pulsar 
magnetospheres may result in quite different emission spectra of 
the magnetospheric X-ray radiation, and some line features may be 
expected for the NS scenario. These line features could be above 
the observational level of the modern or future X-ray missions 
(e.g. Chandra, XMM).
Unfortunately, no detailed theoretical treatment of such
radiative transfer process in the magnetospheres has been
published so far. Such a detailed study is desirable. 
An important bearing is that, if the theoretical results show 
that the intensity of the line features (e.g. $K_\alpha$ line) 
is above the detection level of some future missions, an either 
positive or negative detection of these lines can provide strong 
evidence against or for the BPCSS hypothesis. At present, no line 
feature has been reported from the known spin-powered X-ray pulsars.

Besides the three criteria listed above, any other distinguishing 
emission properties between the vacuum gap and the SCLF gap may act 
as a criterion for differentiating BPCSSs from neutron stars, as long 
as the binding energy problem can not be solved within the neutron 
star scenario.

\section{Conclusions and Discussions}

The idea that pulsars might be strange stars rather than neutron
stars has long been proposed, but no detailed study on the strange
star model of pulsars has been presented previously. 
In this paper, we have explored the possibility and the corresponding 
implications of the idea that pulsars are born as strange stars rather 
than neutron stars. Our main conclusions can be summarized as follows.

1. The lack of a mechanism producing a successful core-collapse
supernova explosion has been a long-standing problem for
supernova explosion studies. The phase transition from nuclear 
matter to SQM presents a natural way to solve this problem.

2. {\em If pulsars are born as strange stars}, very probably
a thick crust can not be formed after the explosion, nor can it
be formed during the long history of its lifetime, unless the
star experiences some special events or happens to be located
in a superdense environment.

3. The strange stars, when they act as pulsars, may have an
atmosphere with mass $\Delta M \sim 5\times 10^{-15} M_\odot$,
and two completely bare polar caps. We refer such stars to bare
polar cap strange stars (BPCSSs). Pulsars in accreting binaries
and the millisecond pulsars should have thick crusts.

4. A direct consequence of the bare polar caps is that BPCSSs should
have vacuum-type inner gaps above their polar caps, regardless of
the orientation of the magnetic axis with respect to the rotation
axis.

5. The thermal conductivity in the bare quark surface is much
larger than that in the neutron star surface. As a result, BPCSSs
may not have hot polar caps.

We have put this BPCSS picture into test with the current available
pulsar broad band observational data. We hope to emphasize that, with
present data, the idea that pulsars are born as strange stars {\em has 
no strong confliction with the observations}, and it even has {\em an 
advantage} to revive the vacuum gap model, which is almost impossible
in neutron stars. The vacuum gap model has been useful in various 
aspects of radio emission theories, and is supported by some recent 
observations. Detections of hot polar caps from normal isolated pulsars
and line features from the X-ray spectra of some rotation-powered 
pulsars will rule out the BPCSS idea, but negative detections with
sufficient sensitivity can be a support to the BPCSS idea.

One interesting question is that ``can some pulsars be strange stars
while some others be neutron stars?''. People tend to accept either
``all pulsars are neutron stars'' or ``all pulsars are strange stars''
for the sake of beautifulness. Caldwell \& Friedman\cite{cf91} proposed
that if some compact objects are strange stars, essentially all 
``neutron stars'' in the disk of the galaxy should be strange stars
due to the strangeness contamination. They thus object the strange star
idea using the glitching phenomenon observed from some pulsars. However,
with some pulsars showing glitches, there is strong evidence that some 
other compact objects (e.g. the millisecond X-ray pulsar SAX J1808.4-3658) 
are strange stars following the equation-of-state arguments\cite{ldw95lbddv99}.
We thus suggest that the actual distribution of the known pulsars might 
be bimodal. In fact, from the discussions above, we can {\em not} draw 
the conclusion that all the pulsars are strange stars, nor can we say
that they are all definitely neutron stars. The main motivation of
this paper is to pose the possibility of regarding strange stars, 
especially BPCSSs, as another possible candidate for pulsars, 
and our idea is subject to the tests of future observations, with some
possible criteria listed in Sect.5.

All our discussions in this paper are based on the assumption that
SQM is an absolutely stable hadronic state, since we have assumed
the existence of the strange stars. Unfortunately, no hitherto-known
theory can validate or reject this assumption.
In the terrestrial physics, searching for the new state
of strong interaction matter, QGP, is the primary goal
of the relativistic heavy-ion laboratories. Many proposed
QGP signatures have been put forward theoretically and many
experimental data have been analyzed. However, the conclusion
about the discovery of QGP is still ambiguous\cite{muller95}.
The confirmation of the existence of
strange stars in the universe, the significance of which might
be comparable to that of confirming black holes, may bring
profound implications to the fundamental physics, and thus
will remain a hot topic in the new century.

{\em Acknowledgments}:
This work is supported by National Nature Sciences Foundation
of China (19803001 and 19673001), by the Climbing project of
China, and by the Youth Foundation of Peking University.

\end{document}